\documentclass[twoside,12pt]{article}
\usepackage[latin9]{inputenc}

\usepackage{amsmath}
\usepackage{float}
\usepackage{indentfirst}
\usepackage{bm}
\usepackage{epsfig}
\usepackage{graphicx}
\usepackage{svg}

\begin{document}
\bibliographystyle{unsrt}




\begin{center}
\LARGE\bf Electron Dynamics in Neutron Scattering with Hydrogen Atoms
\end{center}

\footnotetext{\hspace*{-.45cm}\footnotesize $^\dag$Corresponding author, E-mail: lbfu@gscaep.ac.cn }

\begin{center}
\rm Mingzhao Xing$^{\rm 1}$ \ and  \ Libin Fu$^{\rm 1\dagger}$
\end{center}

\begin{center}
\begin{footnotesize} \sl
${}^{\rm 1}$ Graduate School of China Academy of Engineering Physics, Beijing 100193, China\\
\end{footnotesize}
\end{center}


\vspace*{2mm}

\begin{center}
\begin{minipage}{15.5cm}
\parindent 20pt\footnotesize
In neutron-proton (n-p) scattering experiments, gas targets have been used to measure scattering length by detecting neutrons and recoil protons. Changes in electron dynamics within the gas target have a negligible effect on dynamics of neutrons and protons. However, electron dynamics are sensitive to the specific form of the n-p interaction during the scattering process, providing additional information to derive parameters in nuclear interaction models. We propose a theoretical approach to obtain these parameters from the momentum spectrum of ionized electrons within a  hydrogen atomic gas target. This approach is based on a three-body scattering involving a neutron, a proton and an electron. We model the n-p interaction as the Yukawa potential and obtain the momentum spectrum of ionized electrons through the solution of the Time-Dependent Schr{\"o}dinger Equation. Electron dynamics exhibit significant differences at various potential parameters. These parameters can be determined by comparing numerical calculations with experimental results. Moreover, this approach offers insights into detecting ultrafast scattering processes.

\end{minipage}
\end{center}

\begin{center}
\begin{minipage}{15.5cm}
\begin{minipage}[t]{2.3cm}{\bf Keywords:}\end{minipage}
\begin{minipage}[t]{13.1cm}
neutron scattering, three-body scattering, electron dynamics, ionized electron, momentum spectrum
\end{minipage}\par\vglue8pt

\end{minipage}
\end{center}

\section{Introduction}

Neutron-proton (n-p) scattering plays a crucial role in unraveling the fundamental processes in nuclear physics, contributing to the study of nuclear forces, the properties of nucleons and the underlying principles of nuclear structure. In previous research \cite{RevModPhys.81.1773,Blatt1949,Noyes1963,Lomon1974,Perez2013}, theoretical models for n-p interactions at various energy scales have successfully explained experimental phenomena. The spin singlet $^1S_0$ and the spin triplet $^3S_1$ n-p scattering exist since nucleons have a spin of 1/2. The Yukawa potential \cite{yukawa1935interaction} is widely used to describe the $^1S_0$ n-p interaction. In the Yukawa picture, the interaction between two nucleons is mediated by the exchange of pions. The force range is one of parameters of the Yukawa potential and its value is related to the mass of pions. The accurate value of the force range can not be determined by a given scattering cross-section with fixed incident energy. Therefore, it is necessary to measure additional information such as the electron dynamics.

The gas target has been used to study the $^1S_0$ n-p scattering by detecting the dynamics of neutrons and protons \cite{huhn2000new,johansson2005forward}. Due to the significantly smaller mass of electrons compared to nucleons, changes in electron dynamics have a negligible effect on dynamics of neutrons and protons. However, electron dynamics are sensitive to the specific form of the n-p interaction during the scattering process. We can utilize a gas target composed of hydrogen atoms to obtain the momentum spectrum of ionized electrons within the target. During the scattering process, neutrons collide with hydrogen atoms, resulting in recoiling protons and ionized electrons. The momentum spectrum of ionized electrons can be precisely measured by the highly efficient cold target recoil ion momentum spectroscopy (COLTRIMS) technique \cite{Doerner2000,Ullrich2003,ullrich1997recoil}. Hence, electron dynamics can offer additional information beyond the scattering cross-section to obtain parameters in nuclear interaction models.

In this work, we employ the Yukawa potential as the $^1S_0$ n-p interaction. We investigate dynamic process of the three-body scattering involving a neutron, a proton and an electron through numerically solving the ordinary differential equation
and the Time-Dependent Schr{\"o}dinger Equation (TDSE) \cite{Bandrauk1991,Dijk2007}. We present the momentum spectrum of ionized electrons at various force ranges of the Yukawa potential. The force range can be determined by comparing numerical results with experimental data. Furthermore, we investigated the change in electronic states in the head on n-p scattering process. The ionization probability reaches its maximum when the distance between the two nuclei is nearly at its minimum. By calculating the probability of the ionization and the excitation at various force ranges, we find that the electron is in the ground state or continuum state.

The rest of this paper is organized as follows. In section 2, we introduce
the calculation process of the three-body scattering in detail. In
section 3, we present the numerical results and discussions. A summary
is given in section 4.

\section{Model and method}

In our theoretical model, the three-body system consists of a neutron, a proton and an electron. There are only two types of interactions in this system: the n-p interaction and the Coulomb interaction bwtween the proton and the electron. Masses of the neutron, the proton and the electron are denoted as $m_{n}$, $m_{p}$ and $m_{e}$, respectively. Coordinates of the neutrons, the protons and the electrons are represented by $\boldsymbol{r}_{n}$, $\boldsymbol{r}_{p}$ and $\boldsymbol{r}_{e}$, respectively. The Hamiltonian of this system in the atomic unit is written as
\begin{equation}
H=-\frac{1}{2m_{n}}\Delta_{\boldsymbol{r}_{n}}-\frac{1}{2m_{p}}\Delta_{\boldsymbol{r}_{p}}-\frac{1}{2m_{e}}\Delta_{\boldsymbol{r}_{e}}+V_{N}\left(\boldsymbol{r}_{n}-\boldsymbol{r}_{p}\right)-\frac{1}{\left|\boldsymbol{r}_{e}-\boldsymbol{r}_{p}\right|},
\end{equation}
where $\Delta$ represents the the Laplace operator and $V_{N}$ denotes
the n-p interaction. The last term is the Coulomb potential.

\begin{figure}[H]
\centering{}
\includegraphics[width=8cm]{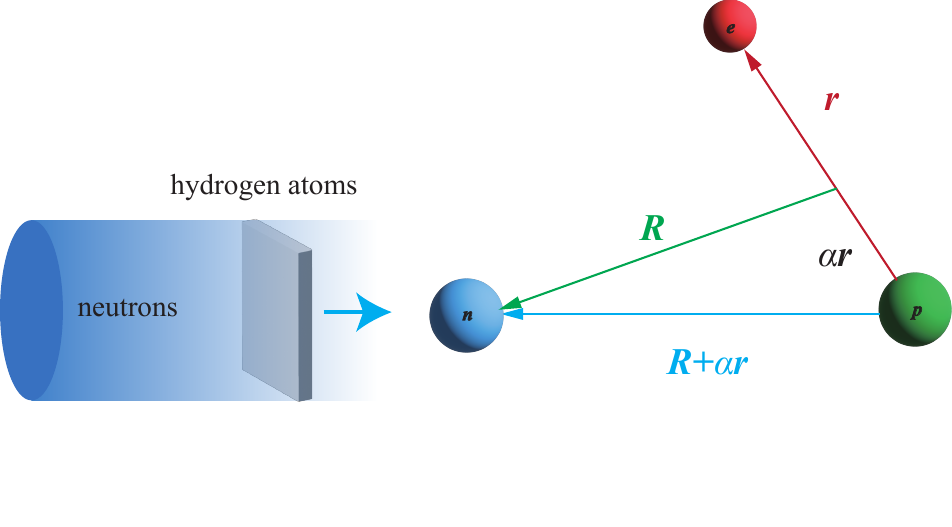}\caption{\label{fig:Schematic-diagram-of}Schematic illustration of the envisaged experiment: an atomic hydrogen gas target is exposed to a neutron beam. The blue (green, red) sphere represents the neutron (the proton, the electron). $\boldsymbol{r}$ denotes the relative position vector from the proton to the electron. $\boldsymbol{R}$ denotes the relative position vector from the center of mass of the hydrogen atom to the neutron. $\alpha$ represents the ratio of the electron's mass to that of the hydrogen atom, which is equal to $1/1837$.}
\end{figure}

The laboratory coordinates $\left(\boldsymbol{r}_{n},\boldsymbol{r}_{p},\boldsymbol{r}_{e}\right)$
can be transformed into the Jacobi coordinates \cite{brauner1989triply} $\left(\boldsymbol{r},\boldsymbol{R},\boldsymbol{R}_{c}\right)$. Here, $\boldsymbol{r}$, $\boldsymbol{R}$, and $\boldsymbol{R}_{c}$ represent vectors denoting the relative distances between the electron and proton, the neutron and the center of mass of the hydrogen atom, and the center of mass of the three-body system, respectively. Then the transformation can be shown as

\begin{align}
\boldsymbol{r} & =\boldsymbol{r}_{e}-\boldsymbol{r}_{p},\\
\boldsymbol{R} & =\boldsymbol{r}_{n}-\boldsymbol{r}_{p}-\alpha\boldsymbol{r},\\
\boldsymbol{R}_{c} & =\frac{m_{n}\boldsymbol{r}_{n}+m_{p}\boldsymbol{r}_{p}+m_{e}\boldsymbol{r}_{e}}{M},
\end{align}
where $M$ is the total mass of the three-body system and expressed as $M=m_{n}+m_{p}+m_{e}$. The symbol $\alpha$ is the ratio of the electron's mass to that of the hydrogen atom, given by $\alpha=m_{e}/\left(m_{p}+m_{e}\right)=1/1837$. In Fig. \ref{fig:Schematic-diagram-of}, the schematic diagram shows coordinates in the three-body system. The blue (green, red) circle represents the electron (the proton, the neutron).

In the Jacobi coordinates $\left(\boldsymbol{r},\boldsymbol{R},\boldsymbol{R_{C}}\right)$,
the Hamiltonian is rewritten as 
\begin{equation}
H=-\frac{1}{2M}\Delta_{\boldsymbol{R}_{c}}-\frac{1}{2\mu_{N}}\Delta_{\boldsymbol{R}}-\frac{1}{2\mu_{e}}\Delta_{\boldsymbol{r}}-\frac{1}{r}+V_{N}\left(\boldsymbol{R}+\alpha\boldsymbol{r}\right),
\end{equation}
where $\mu_{N}$ is the reduced mass of the neutron and hydrogen atom,
$\mu_{e}$ is the reduced mass of the proton and the electron. $\mu_{N}$
and $\mu_{e}$ are represented as follows:

\begin{equation}
\mu_{N}=\frac{m_{n}\left(m_{p}+m_{e}\right)}{m_{n}+m_{p}+m_{e}},
\end{equation}
\begin{equation}
\mu_{e}=\frac{m_{p}m_{e}}{m_{p}+m_{e}}.
\end{equation}

Here, we select the the center-of-mass (CM) coordinate system, where
$\boldsymbol{R}_{c}=0$ and $\Delta_{\boldsymbol{R}_{c}}=0$. Then,
the Hamiltonian $H$ is written as 
\begin{equation}
H=-\frac{1}{2\mu_{N}}\Delta_{\boldsymbol{R}}-\frac{1}{2\mu_{e}}\Delta_{\boldsymbol{r}}-\frac{1}{r}+V_{N}\left(\boldsymbol{R}+\alpha\boldsymbol{r}\right).
\end{equation}
As $\boldsymbol{R}$ becomes larger, $V_{N}$ approaches zero. Hence
the initial state of the three-body system $\Psi\left(\boldsymbol{r},\boldsymbol{R}\right)$
is approximately as follows:
\begin{equation}
\Psi\left(\boldsymbol{r},\boldsymbol{R}\right)=\frac{1}{\left(2\pi\right)^{3/2}}\varphi_{0}\left(\boldsymbol{r}\right)\exp\left(i\boldsymbol{P}_{N}\cdot\boldsymbol{R}\right),
\end{equation}
where $\varphi_{0}$ is the ground state of the electron in the hydrogen
atom and $\boldsymbol{P}_{N}$ is the momentum of the reduced nucleus
(the neutron and the proton).

To elucidate this scattering process, a semi-classical approach is
adopted. Newton's second law is employed to solve for the nucleus's
motion, while the Schr{\"o}dinger equation is used to describe the electron's
motion. In the CM coordinate system, the potential $V_{N}$ depends
on both $\boldsymbol{R}$ and $\boldsymbol{r}$. However, due to the
significantly smaller reduced mass of the proton and the electron
compared to that of the three-body system $\left(\mu_{e}/\mu_{N}\approx1/918\right)$
, the electron's motion has a negligible influence on the nucleus's movement. Thus, in terms of the nucleus, $V_{N}$ can be considered solely as a function of $\boldsymbol{R}$ and denoted as $V_{N}\left(\boldsymbol{R}\right)$.
The classical movement of the nucleus can be expressed via the Newton's second law as

\begin{equation}
\mu_{N}\ddot{\boldsymbol{R}}=-\frac{\partial}{\partial\boldsymbol{R}}V_{N}\left(\boldsymbol{R}\right).
\end{equation}
Then the value of $\boldsymbol{R}$ at time $t$ can be determined.
Concerning the electron, the position of the nucleus has a significant
impact on the electron, resulting that the electron's wave function
is dependent on the nucleus's position. Denoting $\psi\left(\boldsymbol{r},t\right)$ as the electron's wave function, the TDSE for the electron is expressed as follows:
\begin{equation}
i\hbar\frac{\partial}{\partial t}\psi\left(\boldsymbol{r},t\right)=\left[-\frac{1}{2\mu_{e}}\Delta_{\boldsymbol{r}}-\frac{1}{r}+V_{N}\left(\boldsymbol{R}+\alpha\boldsymbol{r}\right)\right]\psi\left(\boldsymbol{r},t\right).\label{eq:TDSE}
\end{equation}

In low-energy (below 5 MeV) n-p scattering process \cite{Noyes1963}, the
singlet state with angular momentum $l=0$ overwhelmingly contributes
to the total cross-section. Within this singlet state, the potential
$V_{N}$ is described using the Yukawa potential model. The Yukawa
potential is related with two parameters, $\lambda$ and $V_{0}$,
which is expressed as \cite{yukawa1935interaction} 
\begin{equation}
V_{N}\left(\boldsymbol{R}+\alpha\boldsymbol{r}\right)=V_{0}\frac{e^{-\left|\boldsymbol{R}+\alpha\boldsymbol{r}\right|/\lambda}}{\left|\boldsymbol{R}+\alpha\boldsymbol{r}\right|},\label{eq:Yukawa potential}
\end{equation}
where $\lambda$ represents the force range, and $V_{0}$ serves
as an integral constant that characterizes the field strength. When
$V_{0}>0$, $V_{N}$ acts as a repulsive potential. When $V_{0}<0$,
$V_{N}$ becomes an attractive potential. In the previous work, the total cross-section $\sigma$ for the Yukawa
potential under the Born approximation \cite{dalitz1951higher} is given by 
\begin{equation}
\sigma=\frac{16\pi\mu^{2}V_{0}^{2}\lambda^{4}}{\hbar^{4}\left(1+4k^{2}\lambda^{2}\right)}\label{eq:cross-section}
\end{equation}
where $\mu$ is the reduced mass of the proton and the neutron, i.e.,
$\mu=m_{n}m_{p}/\left(m_{n}+m_{p}\right)$. $k$ represents the relative
incident momentum between the neutron and the proton. The cross-section
can be obtained through experimental measurements. Then $\lambda$
and $V_{0}$ can be derived according to Eq. (\ref{eq:cross-section}).

The ionized wave function
of the electron is given by:

\begin{equation}
\psi_{ion}\left(\boldsymbol{r},t\right)=\psi\left(\boldsymbol{r},t\right)-\sum_{i}\left\langle \varphi_{i}\mid\psi\right\rangle \varphi_{i}\left(\boldsymbol{r}\right),
\end{equation}
where $\varphi_{i}$ represents the $i$-th ($i=0,1,2...$) bound
state of the electron in the hydrogen atom. Through the Fourier transform,
the ionized wave function of the electron in the momentum representation
is 
\begin{equation}
\phi_{ion}\left(\boldsymbol{p},t\right)=\frac{1}{\left(2\pi\right)^{3}}\int\psi_{ion}\left(\boldsymbol{r},t\right)e^{-i\boldsymbol{p}\cdot\boldsymbol{r}/\hbar}\mathop{\mathrm{d}\boldsymbol{r}.}\label{eq:ionized wave function}
\end{equation}

Here, we numerically calculate the electron's wave function, as shown in Eq. (\ref{eq:TDSE}), and subsequently obtain
the ionization rate and the momentum spectrum of ionized electrons according
to Eq. (\ref{eq:ionized wave function}).

\section{Results and discussions}
\subsection{Ionized electron momentum spectrum}

Due to the rotational symmetry around the neutron incident direction,
our numerical computations are conducted within a two-dimensional
framework. We designate the positive x-axis direction as the orientation
of the incident momentum. The incident momentum is denoted as
$\boldsymbol{P}_{N}=\left(P_{N},0\right)$. The impact parameter is
represented as $b$. We regard the time when the neutron is just affected
by the Yukawa potential as the initial time. At this time, the position
of the neutron is $\boldsymbol{R}=\left(-\sqrt{R_{0}^{2}-b^{2}},b\right)$,
where $R_{0}$ corresponds to the spatial boundary of the potential
$V_{N}$. The range of the impact parameter is $-R_{0}\leq b\leq R_{0}$.
The initial state of the electron is its ground state. We obtain the electron's wave function at time $t$ through TDSE. The symbol $\Delta b$
represents the interval between the impact parameter $b_{i}$ and
$b_{i+1}$. The ionized electron momentum spectrum at a given impact
parameter $b_{i}$ is denoted as $f_{\boldsymbol{p}}\left(b_{i}\right)$.
We then aggregate the momentum spectra corresponding to various impact
parameters, and the total ionized electron momentum spectrum can be
expressed as follows:
\begin{equation}
f_{total}=\frac{2\Delta b}{R_{0}^{2}}\sum_{i}b_{i}f_{\boldsymbol{p}}\left(b_{i}\right).
\end{equation}

\begin{figure}[H]
\begin{centering}
\includegraphics[width=15cm]{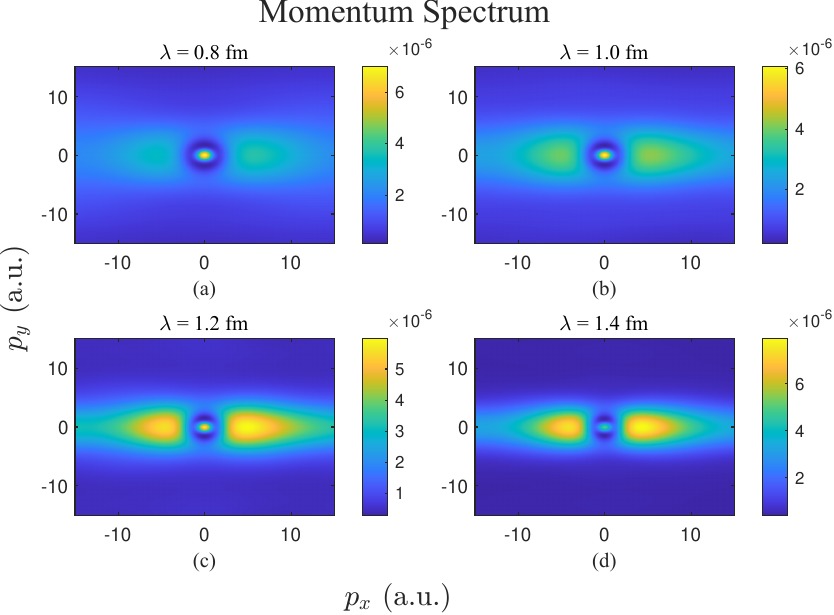}
\par\end{centering}
\begin{centering}
\includegraphics[width=7cm]{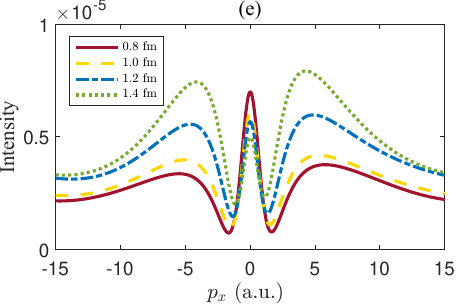}\includegraphics[width=7cm]{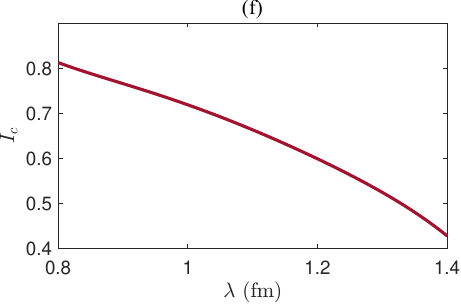}
\par\end{centering}

\caption{\label{fig:Momentum-spectrum}The momentum spectrum of ionized electrons. The effective incident
energy fixed at 1.078 MeV. (a)-(d) correspond to the cases $\lambda$= 0.8 fm,\;1.0 fm,\;1.2 fm
and 1.4 fm respectively. The x-axis and the y-axis correspond to
the momentum along the incident direction and perpendicular to the
incident direction. The color bar represents the value of $f_{total}$. (e)
shows the momentum intensity distribution along the x direction with
$P_{y}=0$. The red, yellow, blue and green lines correspond to $\lambda$=0.8 fm,\;1.0 fm,\;1.2 fm
and 1.4 fm respectively. (f) depicts the variation of the contrast value, denoted as $I_c$, as a function of the force range $\lambda$. $I_c$ illustrates the contrast between the maximum and minimum values of momentum distribution intensity near $p_x=0$, as shown in (e).}
\end{figure}

Fig. \ref{fig:Momentum-spectrum}(a)-(d) illustrate the momentum distribution
of ionized electrons in the final scattering state for various force
ranges $\lambda$=0.8 fm,\;1.0 fm,\;1.2 fm and 1.4 fm through our semi-classical
method. Since the momentum of ionized electrons is dependent on the
incident neutron's effective kinetic energy and the force range $\lambda$,
we have fixed the incident effective kinetic energy at 1.078 MeV \cite{Noyes1963}.
The x-axis and the y-axis in Fig. \ref{fig:Momentum-spectrum}(a)-(d) correspond to the momentum along the incident
direction and perpendicular to it, respectively. The color bar represents
the value of $f_{total}$.The lower limit of the force range ($\lambda$=0.8 fm)
is roughly estimated from the proton's radius. The upper limit of
the force range is estimated based on the mass of the pion: $\lambda_{max}=\hbar/\left(m_{\pi}c\right)\approx$1.4 fm,
where $c$ is the speed of light and $m_{\pi}$ is the mass of the
pion. A notable distinction emerges in the momentum distribution with
different $\lambda$: for smaller $\lambda$ values, ionized electrons
tends to cluster around $\boldsymbol{p}=0$, while higher $\lambda$
values cause the momentum distribution of ionized electrons to broaden
towards both positive and negative directions along the x-axis. To
elucidate the differences between the momentum distribution with various
force ranges, Fig. \ref{fig:Momentum-spectrum} (e) displays the momentum
intensity distribution along the x direction with $P_{y}=0$. As momentum increases, the probability of ionized electrons decreases and then rebounds. Smaller values of $\lambda$
lead to higher momentum intensity around $P_{x}=0$ and fewer amount of rebound.

The contrast value $I_c$ in Fig. \ref{fig:Momentum-spectrum}(f) is defined as
\begin{equation}
I_c=\frac{I_{max}-I_{min}}{I_{max}+I_{min}},\label{I_c}
\end{equation}
where $I_{max}$ and $I_{min}$ represent  the maximum and minimum values of momentum distribution intensity near $p_x=0$ in Fig. \ref{fig:Momentum-spectrum}(e) , respectively. The difference in contrast values can be used to determine the force range of the Yukawa potential by comparing numerical calculations with experimental results.

\subsection{Temporal evolution of electronic states}
In previous studies of scattering problems, the scattering time is typically neglected. The S-matrix element only involves the initial state, the final state and the interaction. In our approach, the time taken by the reduced system, with the reduced mass $\mu_{N}$, to traverse the potential range $V_{N}$ is defined as the scattering time. We can determine the electronic state of the system at each time $t$ during the scattering process. This allows us to establish a relationship
between the ionization probability and the scattering time, providing insights into how the ionization probability evolves throughout the scattering event.

\begin{figure}[h]
\centering{}\includegraphics[width=8cm]{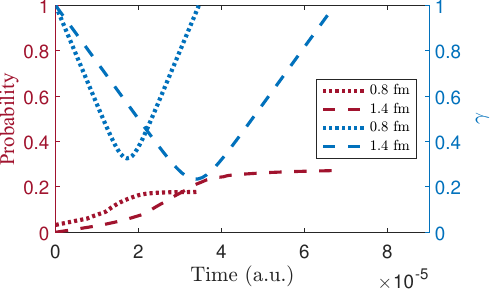}\caption{\label{fig:TDSE} The red lines show the ionization probability as
a function of the the scattering time. The blue lines depict the temporal
evolution of the ratio of relative distance between two nuclei to
the maximum distance of potential energy, $\gamma\left(t\right)=R\left(t\right)/R_{0}$.
the temporal evolution of the ratio of nuclear relative distance to
the maximum potential energy range, denoted as $\gamma\left(t\right)=R\left(t\right)/R_{0}$. The
minimum value of $\gamma$ represents the smallest nuclear relative
distance. The red lines and blue lines represent the variation of ionization probability and $\gamma(t)$
with time, respectively. The dotted lines and dashed lines represent $\lambda$
value of 0.8 fm and 1.4 fm, respectively. The horizontal coordinates
corresponding to the endpoints of each line indicate the total scattering
time.}
\end{figure}

The red lines in Fig. \ref{fig:TDSE} represent the ionization probability
as a function of the the scattering time. The blue lines show the
temporal evolution of the ratio of relative distance between two nuclei
to the maximum distance of potential energy, denoted as $\gamma\left(t\right)=R\left(t\right)/R_{0}$. We
set the impact parameter $b=0$ for simplicity. The dotted lines and
the dashed lines correspond to cases of $\lambda$=0.8 fm and $\lambda$=1.4 fm,
respectively. The situation of $\gamma\left(t\right)=1$ means that
there is no interaction between the two nuclei. The ionization rate
is near its maximum when the distance between the two nuclei is at
its minimum. Then the ionization probability stays almost unchanged. 
 Until now, attosecond pulses \cite{bellini1998temporal,schafer1997high,Paul2001,hentschel2001attosecond} have been used to probe the dynamics of electrons within atoms and molecules \cite{krausz2009attosecond,borrego2022attosecond}. The time duration of the scattering process is less than one zeptosecond according to the results from Fig. \ref{fig:TDSE}, which means observing the nuclear scattering process would require lasers even faster than zeptosecond timescales. Through this theoretical approach, it might be possible to indirectly access the ultrafast nuclear scattering process by probing dynamics of electrons. 

\begin{figure}[h]
\centering{}\includegraphics[width=8cm]{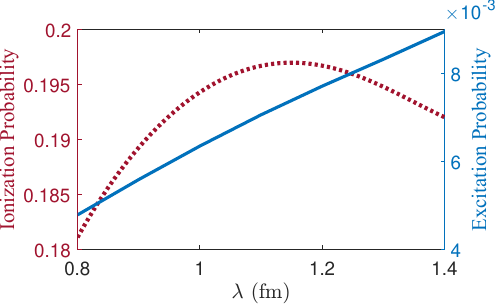}\caption{\label{fig:Ionization Probability}The red dotted line shows the ionization probability
of electrons as a function of the force range $\lambda$. The blue solid line depicts the excitation probability of electrons varying with the force ranges
$\lambda$.}
\end{figure}

To illustrate the influence of the force range $\lambda$ on the electronic state, Fig. \ref{fig:Ionization Probability} depicts the ionization probability (the red dotted curve) and the excitation probability (the blue solid curve) of scattering electrons as a function of the force range $\lambda$, respectively. In this figure, as the force range $\lambda$ decreases, both the total ionization probability and the excitation probability decrease. However, the change in the excitation probability is approximately two orders of magnitude smaller than that in the ionization probability. The impact of the force range on the excitation probability is negligible compared to that on the ionization probability. Thus, the scattering process primarily involves ground states and continuum states of the hydrogen atom.

\section{Conclusion}

In this article we focus on the three-body scattering process involving a neutron, a proton and an electron. Neutrons collide with protons, resulting in recoiling protons and ionized electrons. Electron dynamics are sensitive to the specific form of the neutron-proton interaction during the scattering process, offering additional information beyond scattering cross-section to determine parameters in nuclear interaction models. We report calculations of this process through numerically solving the ordinary differential equation and the Time-Dependent Schr{\"o}dinger Equation. We find momentum spectra of ionized electrons exhibit significant differences at various Yukawa potential parameters. Therefore, these parameters can be determined by comparing numerical calculations with experimental results. 

Numerical results also show the temporal evolution of electronic states during the scattering process. Detailed analyses are made on factors that affect the ionization probability, including the scattering time and the relative distance between two nucleons. Additional discussions have been made on the ionization probability and excitation probability at various Yukawa potential parameters. Our theoretical approach should be useful for further experiments or theoretical analyses.

\section*{Acknowledgements}
This work was supported by the National Natural Science Foundation of China (Grants No. 12088101 and No. U2330401).

\bibliography{npscattering}

\vspace*{-1mm}
\begin{small}\baselineskip=10pt\itemsep-2pt

\end{small}

\end{document}